  \documentclass[twocolumn,prl,aps]{revtex4}

 \usepackage[dvips]{graphicx}
 \usepackage[dvips]{graphics}

\newlength{\bxwidth}\bxwidth=0.8\textwidth

\begin{document}
\title{Dispersion of Magnetic Excitations in Superconducting Optimally Doped ${YBa_{2}Cu_{3}O_{6.95}}$}
\author{D. Reznik$^{1,2\ast}$, P.~Bourges$^2$, L. Pintschovius$^1$, Y. Endoh$^{3,4}$, Y. ~Sidis$^2$,
T. Matsui$^5$, and S. Tajima$^5$.}

\affiliation{$^1$ Forschungszentrum Karlsruhe, Institut f\"ur Festk\"orperphysik, Postfach 3640, D-76021 Karlsruhe, Germany.\\
$^2$ Laboratoire L\'eon Brillouin, CEA-CNRS, CE-Saclay, 91191 Gif sur Yvette, France.\\
$^3$ Institute for Material Research, Tohoku University, Katahira, Aoba-ku, Sendai, 980-8577, Japan.\\
$^4$ International Institute for Advanced Studies, Kizugawadai, Kyoto, 619-0225, Japan.\\
$^5$ Superconductivity Research Laboratory, ISTEC, Shinonome, Koutu-ku, Tokyo, 135-0062, Japan.
}

%  25.40.Fq Inelastic neutron scattering
% 74.72.Bk Superconducting materials Y-based cuprates
% 74.25.Ha Superconductivity Magnetic properties

\begin{abstract}

Detailed neutron scattering measurements of
${YBa_{2}Cu_{3}O_{6.95}}$ found that the resonance peak and incommensurate magnetic scattering induced by superconductivity represent the same physical phenomenon: two dispersive branches that converge near 41 meV and the in-plane wave-vector
\textbf{q}$_{af}$=($\pi$/a, $\pi$/a) to
form the resonance peak. One branch
has a circular symmetry around \textbf{q}$_{af}$ and quadratic downward dispersion from $\approx$41 meV to
the spin gap of 33$\pm$1meV. The other, of lower intensity,
disperses from $\approx$41 meV to at least 55 meV. Our results exclude a quartet of vertical incommensurate rods in \textbf{q}-$\omega$ space expected from spin waves produced by dynamical charge stripes as an origin of the observed incommensurate scattering in optimally-doped YBCO.

\end{abstract}

\maketitle

The magnetic resonance peak observed by inelastic neutron
scattering (INS) is one of the most striking features of
high-T$_c$ superconductors [1-7]. It corresponds to an unusual
enhancement of spin fluctuations in the SC state at the planar
antiferromagnetic (AF) wave vector \textbf{q}$_{af}$ = ($\pi$/a,
$\pi$/a) at an energy E$_r$, which is found experimentally to
scale with T$_c$ as a function of hole doping. First discovered in
optimally doped $\rm YBa_{2}Cu_3O_{6+x}$ \cite{rossat}, its
observation has then been extended to other systems such as ${\rm
Tl_2Ba_2CuO_{6+x}}$ \cite{he02} and ${\rm
Bi_2Sr_2CaCu_2O_{8+x}}$ \cite{fong99}. The role of the magnetic resonance in measured changes of the
fermionic properties of cuprates below T$_c$ is hotly debated
\cite{abanov02,eschrig03,kee02}.

In underdoped $\rm YBa_{2}Cu_3O_{6+x}$ (x=0.6
\cite{mook98}, x=0.7 \cite{fong00,arai99}, x=0.85
\cite{bourges00}), INS measurements revealed additional
incommensurate (IC) spin fluctuations at energies below E$_r$ at
low temperature. They may appear as reminiscent \cite{mook98} of a
quartet of peaks at planar wave vectors \textbf{q}$_{ab}$ =
($\pi$/a(1$\pm\delta$), $\pi$/a) and ($\pi$/a,
$\pi$/a(1$\pm\delta$)) observed in both the SC and normal state of
$\rm La_{2-x}Sr_xCuO_4$ \cite{cheong91}. These are sometimes
interpreted as spin waves originating from stripe fluctuations (fluctuating hole poor AF domains in anti-phase 
separated by fluctuating lines of holes running along a* or b* \cite{tranquada96}).  
However in
weakly underdoped $\rm YBa_{2}Cu_3O_{6.85}$ \cite{bourges00}, both
energy and temperature dependencies of the incommensurability
indicate that IC fluctuations and the magnetic resonance peak are
part of a downward dispersing magnetic collective mode that
exists in the SC state only. Presently, the origin of the IC
magnetic fluctuations, their interplay with the magnetic resonance
peak and their evolution with hole doping are still controversial
issues.

The current study of $\rm YBa_{2}Cu_3O_{6.95}$ focuses on the
search for magnetic signatures of dynamical stripes and on the
dispersion of the magnetic excitations at optimum doping. Throughout
this article the wavevector \textbf{Q} (H,K,L) is indexed in reciprocal lattice
units ($2\pi/a, 2\pi/b, 2\pi/c$).

All INS experiments were performed on the 1T double focusing
triple-axis spectrometer at the ORPHEE reactor. The 1.5 cm$^3$
sample with mosaic spread of 2.2$^{\circ}$ was made up of 3
coaligned high quality twinned single crystals of $\rm
YBa_{2}Cu_3O_{6+x}$ (T$_c$=93K). Lattice constants precisely
measured by high resolution neutron diffraction were: a=3.816 \AA,
b=3.886 \AA, c=11.682 \AA. By comparison with Ref.
\cite{jorgensen90}, the unit cell volume as well as the a-b
anisotropy indicate x=1, the c-axis lattice constant gives x=0.92,
while the T$_c$ is consistent with x=0.95. A different very
precise study of oxygen content vs. lattice parameters in single
crystals produced higher oxygen contents for the same c-axis
lattice parameters \cite{schweiss}, so we conclude that our sample
had x=0.95$\pm$0.02. The c-axis was aligned close to vertical in
order to access the entire Brillouin zone adjacent to
\textbf{Q}=(3/2, 1/2, 1.7). Relaxed vertical resolution in this
orientation maximizes magnetic intensity, which is broad in the
c-direction. The tilt of the sample changed during the scans to
achieve the desired component of \textbf{q} perpendicular to the
ab-plane (q$_c$). A PG filter minimized contamination from higher
order neutrons. PG002 monochromator and analyzer fixed at the
final energy of 14.8 meV was chosen for the study at 35meV (Figs.
1a and 2c). For the scans between 35 and 55 meV (Fig. 2a) we used
a Cu111 monochromator and PG002 analyzer fixed at 30.5 meV. We
also found an experimental condition that dramatically
improved spectrometer resolution over all previous measurements
(energy resolution of $\approx$2.2 meV FWHM; the longitudinal and
transverse q-resolution of $\approx$0.11 \AA$^{-1}$ and
$\approx$0.2 \AA$^{-1}$ respectively) and allowed looking at the
resonance peak in a "magnifying glass". (Fig. 2b) Here we used a
Cu220 monochromator and a PG002 analyzer fixed at 15.2 meV. At energies far above the resonance peak only a
limited part of \textbf{q}-space was measured because the kinematics of the neutron allow access to only a small part of the Brillouin zone. Using an unpolarized neutron beam, we relied on the standard
technique of using temperature dependence to distinguish magnetic
and nuclear scattering \cite{fong95, bourges96}.

\begin{figure}[t]
\centerline{\includegraphics[width= 9 cm]{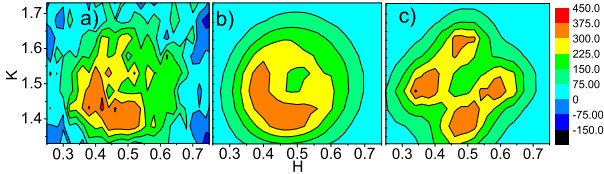}}
%\centerline{\includegraphics[width=\textwidth]{fig1.pdf}}
\caption{ {\label{fig1}} a) 10K intensity minus 100K intensity at
35 meV (same data as in figure 2c) b) Model calculations
(represented in figure 2c by blue lines) The apparent lack of
symmetry around \textbf{q}$_{af}$ after convoluting with the
resolution function is a focusing effect illustrated in figure 3b,
which is a clear signature of dispersion. c) Predictions of the stripe model. The small differences in intensities of the four spots are due to the magnetic form factor.  Magnetic correlation length used in the calculations was the same as in (b) and (c)(see text). }
\end{figure}

Constant energy scans taken at E=41meV and T=100K showed only a flat background in agreement with previous results for this composition \cite{fong95,bourges96}. However, rather strong features with \textbf{q}-dependence expected from magnetic scattering persisted to 100K at E=44meV and above. Further heating to 300K did not make them disappear entirely. Therefore, their magnetic origin needs to be confirmed by further measurements. In the following, our analysis is based on intensity differences between 10K and 100K, which are confidently assigned to magnetic scattering.

In order to search for evidence of dynamical stripes in optimally-doped YBCO we mapped in \textbf{Q}-space the scattering intensity at 35 meV, which is just above the measured spin gap of 33$\pm$0.5meV (data not shown in figures)
and well below E$_r$ ($\approx$41meV). High resolution scans at 10K and 100K shown in figure 2f clearly show the extra IC magnetic scattering appearing at 10K on top of a broad feature originating from phonons. The contour plot of the
intensity difference between 10K and 100K at 35 meV (Fig. 1a) measured with lower resolution also
clearly shows IC magnetic signal. However, the data show strong asymmetry around \textbf{q}$_{af}$, a high symmetry point of the nearly tetragonal YBCO reciprocal lattice. This apparent anomaly can be easily
explained by the tilt of the resolution ellipsoid in \textbf{q}-$\omega$ space that breaks the symmetry in the experiment and results in well-understood focusing/defocusing effects. In fact a spectral function S(\textbf{q},$\omega$) that is dispersive and circularly symmetric around \textbf{q}$_{af}$ results in an intensity distribution that closely matches the data (Figs. 1b, 2c). On the other hand, the spin wave model for the stripe picture that gives a good description of the magnetic response of stripes in the nickelates \cite{bourges03} should produce a quartet of peaks as calculated in \cite{tranquada04}. After convolution with the calculated spectrometer resolution the model prediction differs from the data far beyond the experimental uncertainty even if a finite correlation length of the stripes is assumed (Fig. 1c). So we can rule out the "spin waves due to stripes" model, though the excitation spectrum of the stripe phase of $\rm La_{1.875}Ba_{0.125}CuO_4$ \cite{tranquada04} is not consistent with such a model either.

\begin{figure*}[ptb]
\begin{center}
%\centerline{\includegraphics[width= cm]{fig2.eps}} \caption{
\includegraphics[width=16.5 cm]{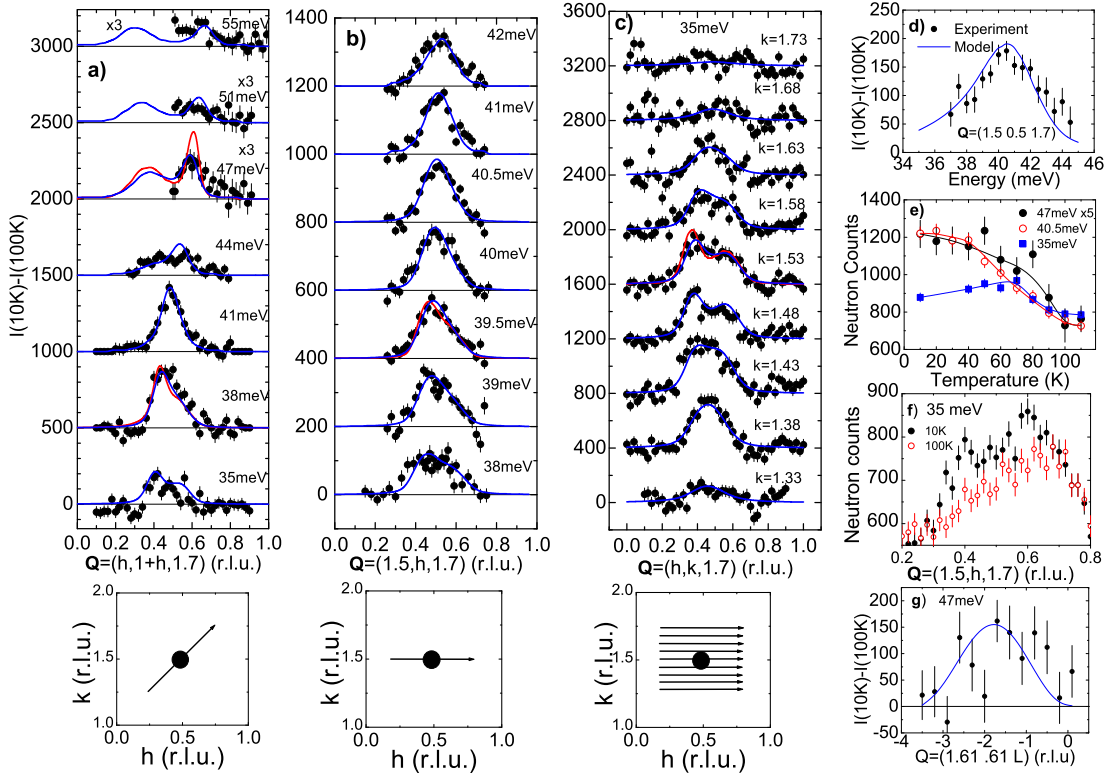} \caption{
{\label{fig2}} a-d) Intensity at 10K minus intensity at 100K (data
points) compared with the model illustrated in Fig. 3 convoluted
with the resolution function. Blue/red lines represent the model
with/without the broadening described in the text. The energy
resolution was a - 5meV, b,d - 2.2 meV, and c - 4meV. A small
linear term was subtracted from the scans in c to make the
background the same on both sides. Overall amplitude differs for
a, b and d, and c, but is the same for each set of calculated
curves. The diagrams below a, b, and c represent the scan directions projected onto the h-k plane for the plots above. e) Scattering intensity at (\textbf{Q},$\omega$)=((1.4 0.4
1.7), 35meV), ((1.5 0.5 1.7), 40.5meV), and ((1.6 0.6 1.7), 47meV)
vs. temperature. Constant intensity was subtracted from 35 meV and
47 meV data to scale together the 10-100K intensities. f) High resolution scans at 35 meV measured with Cu111/PG002 monochromator/analyser and E$_f$=13.4meV at 100K and 10K. g)Intensity at 10K minus intensity at 100K (data
points) compared with the odd bilayer magnetic structure factor (blue line).}
\end{center}
\end{figure*}

Further measurements (Fig. 2a) revealed that the magnetic signal at 35 meV arises from a branch that disperses to lower energies from E$_r$ and cuts off at the spin gap. We also found another branch that disperses upward above E$_r$ persisting above 55 meV. This upper branch has not been previously observed because its peak intensity
at energies where it is well separated from the lower branch is
much smaller than at E$_r$. The raw data at 47meV and 10K (not shown in figures) were consistent
with circular symmetry and not with four IC peaks, though they do not rule out a more complex \textbf{q}-space lineshape. 
Figure 2e shows temperature dependence at three energies with
\textbf{q} fixed at the maximum of the intensity change between 10K
and 100K at each energy. The intensities at 40.5 and 47 meV
increase and then saturate with decreasing temperature below
T$_c$. At 35meV the intensity peaks at $\approx$70K as observed in
$\rm YBa_{2}Cu_3O_{6.85}$ \cite{bourges00} hinting at a nontrivial temperature dependence, which needs to be investigated further. Along L the upper branch has a broad maximum at
L=1.7, thus it is odd in the bilayer as is the
resonance peak (Fig. 2g).

\begin{figure}[t]
\centerline{\includegraphics[width= 9 cm]{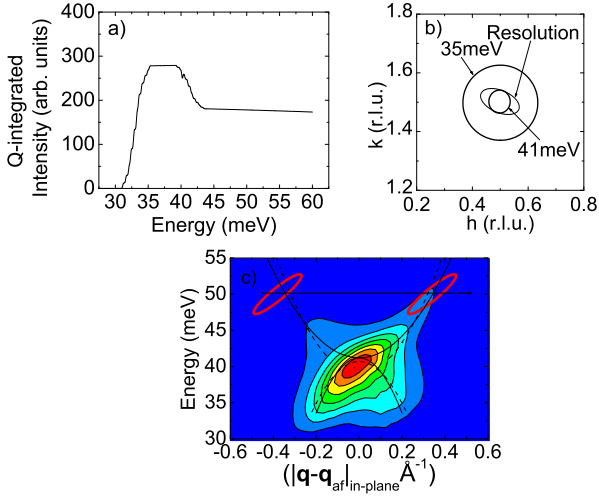}}\caption{
%\centerline{\includegraphics[width=\textwidth]{fig1.pdf}} \caption{}
{\label{fig3}} A model for S(\textbf{q},$\omega$) in good
agreement with most of the data (Fig. 2): a) \textbf{q}-integrated density
of states derived from the model. b)
Schematic of resolution effects in the a-b plane. The oval
represents the a-b plane projection of the resolution ellipsoid,
and the circles represent cuts of S(\textbf{q},$\omega$) at
$\omega$ = 40.5 and 35 meV. The entire S(\textbf{q},$\omega$) fits
into the resolution ellipsoid near 41 meV, which results in
enhanced scattering intensity at the resonance peak energy. c) Solid lines represent the \textbf{q}-$\omega$
dispersion relation. The color plot represents this model convoluted
with experimental resolution. Red lines show the
resolution ellipsoid. Its tilt is responsible for better peak definition on
the left/right side for the upper/lower branch respectively. The dashed line corresponds to a different dispersion relation that also fits the data. }
\end{figure}

We then utilized the high resolution condition to focus on the
dispersion of the two branches where they converge to 41 meV and
\textbf{q}$_{af}$. The constant \textbf{q} scan through
\textbf{q}$_{af}$ (Fig. 2d) shows the well-known resonance peak at
41meV, but the scattering has considerable broadening in
\textbf{q}-space at energies away from 41meV (Fig. 2b).
However the \textbf{q}-integrated
S(\textbf{q},$\omega$), S($\omega$), is roughly the same at all energies in Fig.
2b because of the broadening of
S(\textbf{q},$\omega$) along both H and K.

To get a more quantitative estimate of S($\omega$) we modelled the data (Fig. 2)
with the following dispersions (solid lines in Fig. 3c):

For 33meV$<\omega<\omega_1$: $\omega$ =
-a$\mid\textbf{q}-\textbf{q}_{af}\mid^2$+$\omega_0$;

For $\omega>\omega_2$: $\omega$ =
b$\mid\textbf{q}-\textbf{q}_{af}\mid^2$+$\omega_0$;

where $\omega_{0}$ = 41.2meV, a = 191 meV\AA$^{2}$, b = 75
meV\AA$^{2}$ (Values of b up to $\approx$ 100 give a good fit as well). The low-energy cutoff at 33meV represents
the experimentally determined spin gap. An infinitely sharp ($\delta$ function of the above dispersions) S(\textbf{q},$\omega$) gives a reasonably good agreement with the data (red lines in Fig. 2) in the lower branch, but is too narrow for the upper one. In order to better fit the data, isotropic Lorenzian broadening in \textbf{q} and $\omega$ was
applied to include a finite correlation length of 55\AA\ and an
energy width of 2.2meV FWHM (blue lines in Fig. 2). The ratio of the amplitudes of the upper and lower branch is found to be 0.65 to best agree with the data.

Experimental uncertainty allows slightly different values of $\omega_{0}$, a, and b. There also
may be a gap of less than 1meV between the branches or one or both
branches may not "close" at \textbf{q}$_{af}$ (dashed line in Fig. 3c). 

Fitting the scattering intensity at different energy transfers with the same amplitude places stringent constraints on the dispersion relation exponent. Higher powers than 2 would flatten the dispersion manifold near $\omega_0$ and strongly enhance the predicted scattering intensity there whereas powers lower than 2 would suppress it relative to the expectations of quadratic dispersion. Neither such enhancement or suppression is observed experimentally, which means that the quadratic terms dominate the dispersion relation. This also implies flat S($\omega$) in each branch (Fig. 3a). The important caveat is that the differences between the data and the model at 51 and 55 meV hint at a more complex behavior there. Convergence of S(\textbf{q},$\omega$) to \textbf{q}$_{af}$ at $E_{r}$ (Fig. 3b) can account for all intensity in the resonance peak (Refs. \cite{rossat,fong95}) as demonstrated by the excellent agreement between the model and the data in figure 2d.

Somewhat similar \textbf{q}-$\omega$
dispersions, though with a different temperature dependence, was reported for $\rm
YBa_{2}Cu_3O_{6.7}$ \cite{arai99} and more recently for the stripe phase of $\rm La_{1.875}Ba_{0.125}CuO_4$ \cite{tranquada04}. 

A number of theories predict or assume dispersive magnetic branches in copper oxide superconductors. A 2D metal with strong antiferromagnetic correlations in the d-wave SC state is characterized by dispersive magnetic collective modes \cite{onufrieva02}.  The phenomenological
spin-fermion model assumes an upward-dispersing branch similar
to the observed one though it does not predict the
downward-dispersing one \cite{morr98}. A recent calculation based on the t-t'-J model yielded results somewhat similar to ours \cite{sega03}. An RPA calculation based on the Fermi liquid scenario predicts an incommensurate-commensurate-incommensurate evolution of the magnetic signal with energy \cite{levin00,norman01}. A more
quantitative comparison is required to understand the relevance of
these models. Most importantly, many
theories have to be reevaluated in light of the finding that most
spectral weight of magnetic excitations appears at energies above
and below the resonance peak even at optimum doping.

The crystal growth was supported by the New Energy and Industrial
Technology Development Organization (NEDO) as Collaborative
Research and Development of Fundamental Technologies for
Superconductivity Applications.

\normalsize{$^\ast$To whom correspondence should be addressed;
E-mail: reznik@llb.saclay.cea.fr }

\end{document}